# Thermal diffusivity characterization of impacted composites using evaporative cryocooling excitation and inverse physics-informed neural networks

Pengfei Zhu, *Graduate Student Member, IEEE*, Hai Zhang*, *Member*, *IEEE*, and Stefano Sfarra, Fabrizio Sarasini, Rubén Usamentiaga, *Senior Member, IEEE,* Gunther Steenackers, Clemente Ibarra-Castanedo, and Xavier Maldague, *Senior Member, IEEE*

*Abstract*—The thermal diffusivity measurement of impacted composites using pulsed methods presents an ill-posed inverse problem influenced by multiple factors such as sample thickness, cooling duration, and excitation energy. In this study, a novel excitation method—evaporative cryocooling—was introduced for measuring the thermal diffusivity of tested samples. Compared to conventional excitation modalities, evaporative cryocooling excitation is compact, portable, and low cost. However, evaporative cryocooling cannot be considered a pulsed method due to its prolonged excitation duration. In general, it is difficult to measure thermal diffusivity based on non-impulsive pulsed excitation at times commensurate with the pulse duration, often due to ill-defined pulse shape and width and the subsequent potentially complicated thermal response which may be subject to diffusive broadening. To address this challenge, inverse physics-informed neural networks (IPINNs) were introduced in this work and integrated with an evaporative cryocooling method. The Parker method combined with a photothermal method was employed as a reference. To improve the accuracy of both IPINNs and Parker methods, terahertz time-domain spectroscopy (THz-TDS) was employed for measuring the thickness of impacted composites. Simulations and experimental results demonstrated the feasibility and accuracy of the IPINN-based approach.

This work was supported in part by the Natural Sciences and Engineering Research Council of Canada (NSERC) through the CREATE-oN DuTy! Program under Grant 496439-2017, in part by the Canada Research Chair in Multi-polar Infrared Vision (MIVIM). (Corresponding author: Hai Zhang.)

Pengfei Zhu, Hai Zhang, Clemente Ibarra-Castanedo and Xavier Maldague are with the Department of Electrical and Computer Engineering, Computer Vision and Systems Laboratory (CVSL), Laval University, Québec G1V 0A6, Québec city, Canada (e-mail: pengfei.zhu.1@ulaval.ca; hai.zhang.1@ulaval.ca; clemente.ibarra-castanedo@gel.ulaval.ca; xavier.maldague@gel.ulaval.ca). Hai Zhang is also with the Centre for Composite Materials and Structures (CCMS), Harbin Institute of Technology, Harbin 150001, China (hai.zhang@hit.edu.cn).
Stefano Sfarra is with the Department of Industrial and Information Engineering and Economics (DIIIE), University of L'Aquila, I-67100, L'Aquila, Italy. (e-mail: stefano.sfarra@univaq.it).
Fabrizio Sarasini is with the Department of Chemical Engineering Materials Environment & UDR INSTM, Sapienza University of Rome, Rome, Italy. (e-mail: fabrizio.sarasini@uniroma1.it)
Rubén Usamentiaga is with the Computer Science and Engineering Department at the University of Oviedo (rusamentiaga@uniovi.es).
Gunther Steenackers is with the InViLab Research Group, University of Antwerp, Antwerp, Groenenborgerlaan 171, Belgium (gunther.steenackers@uantwerpen.be)

## I. INTRODUCTION

Thermal diffusivity ($\alpha$) measurements play a key role in the design of instrumentation systems where temperature and thermal stress can complicate signal response [1]-[5]. This is due to the fact that thermal diffusivity controls the internal heat propagation velocity of materials ($q = -k\nabla T$, where $q$ is the heat flux, and $\nabla T$ is the temperature gradient). Local variations of thermal diffusivity are suggested as indicators of severity of impact damage in composite materials [6]. However, complex failure modes (fiber fracture, matrix fracture, delamination) and anisotropic material properties significantly limit the measurement accuracy of thermal diffusivity. Nevertheless, reported research in the literature has ignored the thickness variation after the impact [7]-[11]. Furthermore, conventional methods use a laser or flash lamp as the excitation source. These techniques require costly equipment, including infrared detectors and high-power pulsed lasers (or flash lamps). The entire measurement process must be conducted in specialized research laboratories. Therefore, there is a need to develop novel, portable, and low-cost excitation sources.

State-of-the-art thermal diffusivity measurements can be divided into time-domain and frequency-domain techniques. The time-domain techniques usually depend on transient modalities, such as laser flash methods [12], transient thermal grating (TTG) methods [13], and pulsed photothermal displacement techniques [14]. The laser flash technique (Parker method) was developed in 1961 by Parker et al. [12], and it is commonly used for thermal diffusivity measurements since it is contactless, non-destructive, and highly accurate [15,16]. The laser flash method uses optical heating as an instantaneous heating source to excite the front surface of a sample, and a thermocouple is employed to record the rear surface temperature variation. The thermal diffusivity can be calculated based on a one-dimensional transmission heat conduction model. Transient thermal gratings (TTG) were first introduced by Eichler et al. [13] in 1973, who measured the thermal diffusivity in the in-plane direction of a solid sample. The TTG technique uses the interference of two pulsed laser beams to generate a spatially periodic thermal grating on a sample surface, the relaxation dynamics of which is monitored via a probe beam diffraction and is governed by the material's thermal diffusivity. A typical pulsed photothermal displacement method is the pulsed photothermal mirror proposed by Astrath et al. in 2007 [14]. In this technique, a single pulse heats the sample and causes the subsequent deformation due to thermal expansion. As the sample surface becomes deformed, the probe beam is focused or defocused due to the thermal mirror effect [14,17]. The thermal diffusivity can be calculated by analyzing the change in the intensity profile of the central portion of the beam in the far field.

Frequency-domain techniques include photothermal emission [18], photothermal beam deflection (PBD) [19], photothermal displacement [20], thermal-wave cavity (TWC) photopyroelectric method [21], etc. The photothermal emission method was proposed by Kanstad and Nordal [18] in 1979. In this method, an incident modulated radiation heats periodically the sample surface, and an infrared camera records the periodic temperature fluctuations. By comparing the phase shift between original waveform and recorded signals, the thermal diffusivity can be extracted. Photothermal beam deflection (PBD) was first proposed in 1980 by Boccara et al. [19]. This technique relies on the deflection of a probe beam caused by refractive index gradients in the adjacent gas layer, induced by modulated laser heating, to extract the out-of-plane thermal diffusivity of the sample. Photothermal displacement (PTD) was developed by Olmstead in 1983 [20]. It measures the surface displacement of a sample caused by periodic heating from a modulated pump laser, with the resulting thermal expansion detected via changes in the reflection angle of an obliquely incident probe beam. The thermal-wave cavity (TWC) method was developed by Shen and Mandelis in 1995 [21]. It operates by constructing a resonant thermal-wave cavity, where a thin aluminum foil serves as an intensity-modulated laser-induced thermal-wave oscillator, and a pyroelectric polyvinylidene fluoride (PVDF) film functions as both, the signal transducer and the standing-wave generator. By scanning the modulation frequency, the system exhibits fundamental and higher-order thermal-wave resonances, with overtone amplitudes attenuated due to thermal diffusion characteristics. The TWC technique enables high-precision thermal diffusivity measurements.

Recently, researchers tried to combine deep learning with photothermal techniques for measuring thermal diffusivity. For instance, some researchers use simulation results as training datasets [22]. Then the experimental results were fed into this trained network for prediction. Similarly, other researchers tried to use experimental results to train the networks [23]. However, these methods have low robustness and accuracy.

In this study, we developed a simple yet effective method for measuring out-of-plane thermal diffusivity by integrating inverse physics-informed neural networks (IPINNs) with evaporative cryocooling excitation. This novel approach incorporates the fundamental principles of heat conduction into the neural network training process, thereby reducing over-reliance on purely data-driven models. Simultaneously, IPINNs integrate the heat conduction equation with experimental data to strike a balance between physical modeling and empirical observation. The laser flash method was also employed for comparison with the IPINN approach. The research reported in the literature [6]-[11] did not consider the thickness variation after the impact damage when measuring thermal diffusivity based on the laser flash method. To address this omission, we employed terahertz time-domain spectroscopy (THz-TDS) to accurately measure specimen thickness after impact, thereby improving the precision of thermal diffusivity measurements. Finally, we investigated the thermal diffusivity characteristics of composites subjected to varying impact energies.

## II. Methodology

Following the theoretical and experimental descriptions of the foregoing sections, an inverse physics-informed neural network (IPINN) approach combined with an evaporative cryocooling method for measuring thermal diffusivity was developed. This framework integrates a cooling heat transfer modality with an inverse problem solver (IPINN). To optimize the training datasets for IPINN and improve the accuracy of results obtained compared to the conventional laser flash method, terahertz time-domain spectroscopy (THz-TDS) was used to precisely measure the sample thickness.

### A. Conventional Laser Flash Method

According to the principles of the laser flash method, thermal diffusivity can be determined from the thermal response of the rear face of a sample after the front face is subjected to a laser or flash lamp pulse. Theoretically, the temperature rise on the rear surface as a function of time can be expressed as [12]:

$$T(t) = \frac{Q}{\rho C_p L}[1 + 2\sum_{n=1}^{\infty}(-1)^n \exp(\frac{-n^2\pi^2}{L^2}\alpha t)] \quad (1)$$

where $L$ is the sample thickness, and $Q$ is the radiation energy of the pulse. Eq. (1) can be simplified as:

$$W(t) = 1 + 2\sum_{n=1}^{\infty}(-1)^n \exp[-n^2\eta(t)] \quad (2)$$

where $W(t) = \frac{T(t)}{\frac{Q}{\rho C_p L}} = \frac{T(t)}{T_{max}}$ and $\eta(t) = \frac{\pi^2\alpha t}{L^2}$ are dimensionless parameters, $T_{max}$ is the maximum temperature at the rear side. The thermal diffusivity can be determined by a known specimen thickness $L$ and the time $t_{1/2}$ at which the temperature reaches half the maximum value:

$$\alpha = \frac{1.38L^2}{\pi^2 t_{1/2}} \quad (3)$$

The laser flash method relies on a standard test method E1461 [24] (which outlines the procedure for determining thermal diffusivity) and standard ASTM practice E2585 [25] (which provides practical guidance to complement ASTM E1461, including recommendations for data analysis, test setup optimization, and uncertainty evaluation). Then, this technique was refined using different time parameters, which is called laser flash method. The following equations are introduced to estimate the in-plane diffusivity [26,27]:

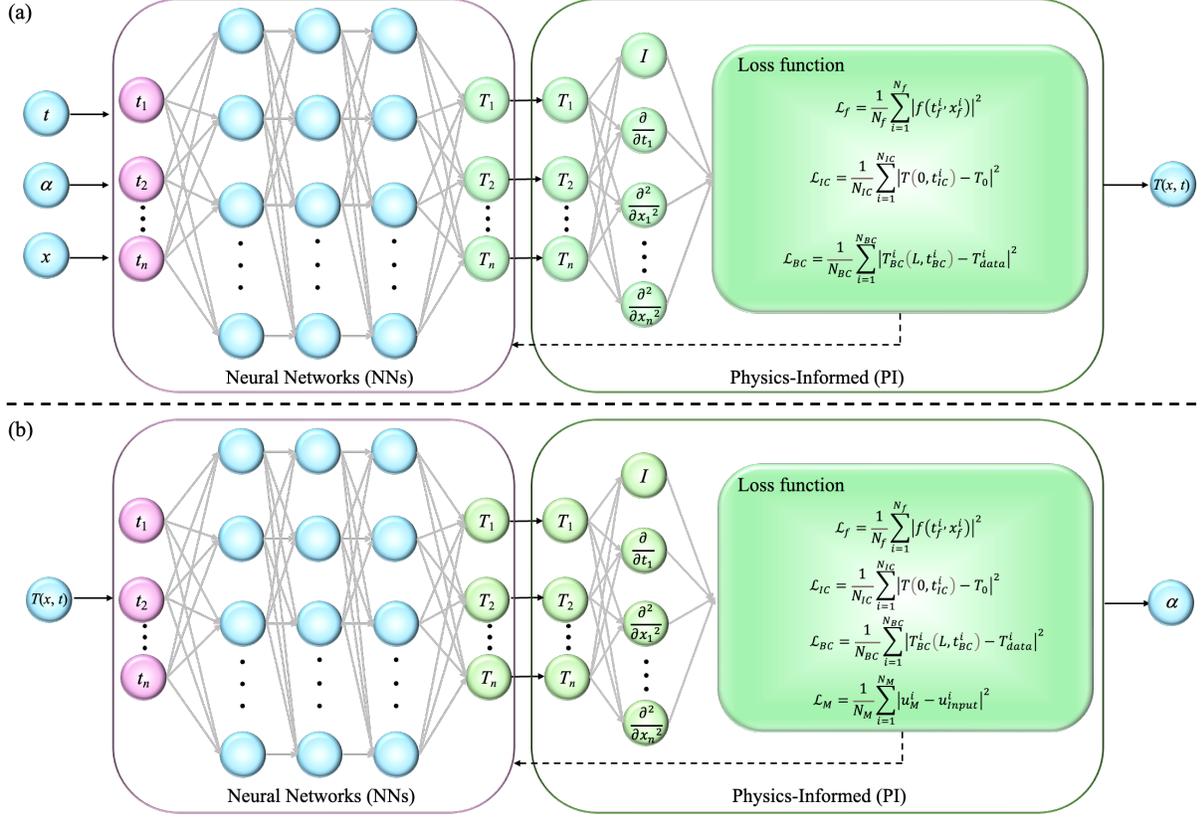

**Fig. 1.** The schematic image of (a) physics-informed neural networks (PINNs) and (b) inverse physics-informed neural networks (IPINNs).

$$\alpha = \frac{L^2}{t_{5/6}}[0.818 - 1.708\frac{t_\frac{1}{3}}{t_\frac{5}{6}} + 0.885(\frac{t_{1/3}}{t_{5/6}})^2] \quad (4)$$

$$\alpha = \frac{L^2}{t_{5/6}}[0.954 - 1.581\frac{t_\frac{1}{2}}{t_\frac{5}{6}} + 0.558(\frac{t_{1/2}}{t_{5/6}})^2] \quad (5)$$

$$\alpha = \frac{L^2}{t_{5/6}}[1.131 - 1.222\frac{t_{2/3}}{t_{5/6}}] \quad (6)$$

where the thermal diffusivity $\alpha$ corresponds to the mean value of the above three results; $L$ is the thickness of the sample; $t_{1/2}$, $t_{1/3}$, $t_{2/3}$, and $t_{5/6}$ correspond to times when temperature equals 1/2, 1/3, 2/3, and 5/6 of its maximum value, respectively. Although the Parker method is based on one-dimensional heat conduction model, it still maintains high measurement accuracy, as validated in numerous studies [28]-[31].

### B. Inverse Physics-Informed Neural Networks

The evaporative cryocooling method cannot be regarded as equivalent to the flash laser method, as it inherently represents a form of long-pulse excitation rather than impulse-response excitation. In general, it is difficult to measure thermal diffusivity based on non-impulsive pulsed excitation at times commensurate with the pulse duration, often due to ill-defined pulse shape and width and the subsequent potentially complicated thermal response which may be subject to diffusive broadening. This can be compounded by the lack of easy-to-follow criteria for assigning boundary conditions such as natural convection in the interfacial heat transfer process, the latter represented by a Grashof number Gr >> 2,000 in case of large thermal gradients between the photothermally excited medium and the surrounding ambient [32]. To address this issue, we propose the use of inverse physics-informed neural networks (IPINNs). A schematic illustration of the proposed method is shown in Fig. 1.

Before introducing IPINNs, it is important to first outline the fundamental principles of PINNs. In general, differential equations play a crucial role in mathematics, physics, and engineering. A standard neural network is typically trained using input-output data pairs, with a loss function defined to minimize the difference between predicted and target outputs. In contrast to conventional neural networks, PINNs incorporate additional loss terms that enforce compliance with underlying differential equation. The one-dimensional heat conduction problem with long-pulse excitation can be formulated as follows:

$$T_t = \alpha_x T_{xx} + \alpha_y T_{yy} + \alpha_z T_{zz} \quad (7)$$

$$T(x,0) = T_0 \quad (8)$$

$$\frac{dT}{dz}\Big|_{(x,y,0,t)} = Q(x,y,0,t) \quad (9)$$

$$\frac{dT}{dt}\Big|_{(i=\Gamma',t)} = h[T(i=\Gamma',t) - T_\infty], \quad i = x, y, \text{and } z \quad (10)$$

where $\alpha_x$, $\alpha_y$, and $\alpha_z$ are the thermal diffusivity along $x$, $y$, $z$ directions, respectively. $\Gamma'$ denotes the external surface except the front surface ($z = 0$). $T_\infty$ is the ambient temperature. In PINNs, the physical residual can be defined as

$$f = T_t - (\alpha_x T_{xx} + \alpha_y T_{yy} + \alpha_z T_{zz}) \quad (11)$$

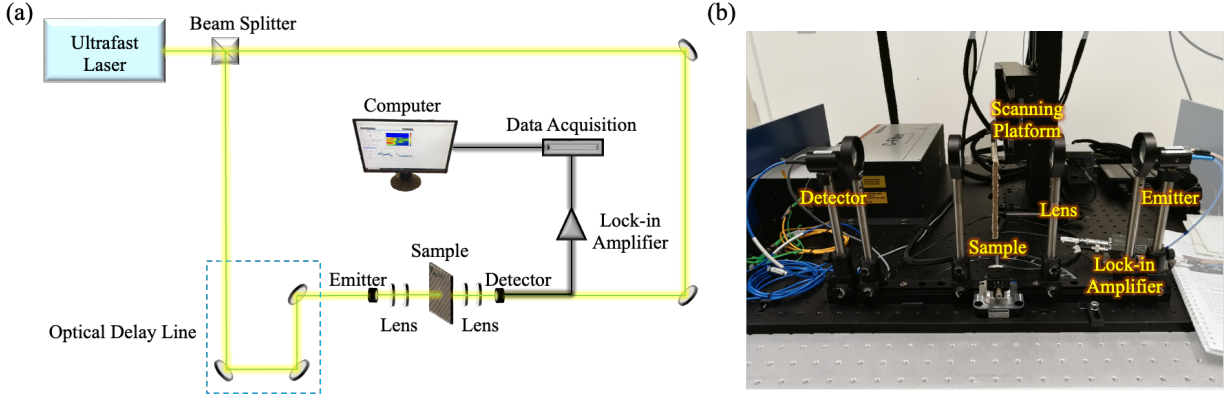

**Fig. 2.** Schematic (a) and experimental (b) setup of the THz-TDS system.

The loss function can be defined as

$$\mathcal{L} = \mathcal{L}_f + \mathcal{L}_{IC} + \mathcal{L}_{BC} \quad (12)$$

$$\mathcal{L}_f = \frac{1}{N_f} \sum_{i=1}^{N_f} |f(t_f^i, x_f^i)|^2 \quad (13)$$

$$\mathcal{L}_{IC} = \frac{1}{N_{IC}} \sum_{i=1}^{N_{IC}} |T(0, t_{IC}^i) - T_0|^2 \quad (14)$$

$$\mathcal{L}_{BC} = \frac{1}{N_{BC}} \sum_{i=1}^{N_{BC}} |T_{BC}^i(L, t_{BC}^i) - T_{data}^i|^2 \quad (15)$$

where $\mathcal{L}_{IC}$ and $\mathcal{L}_{BC}$ denote the loss functions of initial and boundary data, and $\mathcal{L}_f$ enforces the structure imposed by Eq. (12) at a finite set of collocation points.

In PINNs, the inputs typically include the pre-defined time and specimen thickness, while the output is the thermal response. In contrast, IPINNs take time, spatial coordinates, and thermal response as inputs, and the output is the thermal diffusivity. The overall network architecture of IPINNs is similar to that of PINNs. However, in the physical residual term (Eq. (12)) and the loss function (Eq. (13)) of IPINNs, the thermal diffusivity is treated as a variable and is learned during the training process. Furthermore, IPINNs introduce an extra loss component that enforces the predicted thermal response to match the solution of the underlying heat conduction equation:

$$\mathcal{L}_M = \frac{1}{N_M} \sum_{i=1}^{N_M} |u_M^i - u_{input}^i|^2 \quad (16)$$

### C. Terahertz Time-Domain Spectroscopy Technique

Accurate measurement of specimen thickness variation is crucial, especially in cases involving impact-induced damage. This allows us to determine whether changes in thermal diffusivity are attributed to thickness variation or to alterations in material properties. The THz-TDS imaging system is shown in Fig. 2. An ultra-fast laser pulse is split into a pump beam and a reference beam. The pump beam is time-delayed via an optical delay line and directed to a THz emitter, which generates linearly polarized THz radiation. This THz wave passes through the sample and is collected by a detector. The reference beam served as the sampling signal at the detector. The sampled signal is then processed by a lock-in amplifier to enhance weak signals for data acquisition.

The THz system was manufactured by Menlo Systems GmbH, Munich, Germany. It features a frequency resolution of 1.2 GHz and a repetition rate of 100 MHz. The experiments were conducted in transmission mode with a scanning step of 0.5 mm. Additionally, the ambient temperature was controlled at 22 °C ± 0.1 °C, with relative humidity maintained at 50% ± 2 %.

The refractive index $n$ and absorption coefficient $\mu_a$ provide insight into the material's unique microscopic structure, molecular arrangement, and composition [33,34]. A commonly used measurement approach is transmission-mode THz-TDS, as it minimizes surface effects and exhibits lower photonic loss compared to the reflection mode.

The refractive index can be calculated based on the phase difference between sample signal and reference signal [35]:

$$n(\omega) = 1 + \frac{c}{2\pi\omega d}(\varphi_s(\omega) - \varphi_r(\omega)) \quad (17)$$

where $\varphi_s$ and $\varphi_r$ denote the phase angles of the sample signal and reference signal, $c$ is the speed of light, $\omega$ is angular frequency, and $d$ is sample thickness. According to Eq. (17), it is possible to calculate the thickness of the specimen if refractive index is pre-known. However, due to strong interactions between THz photons and individual molecules, the phase velocity of THz beams differs from the group velocity (i.e., the dispersion). Therefore, it is possible to calculate the thickness according to the time delay between sample signal and reference signal:

$$d = \frac{c}{n(\omega)-1}(t_r - t_s) \quad (18)$$

where $t_r$ and $t_s$ denote the peak time of reference signal and sample signal.

### III. EXPERIMENTAL AND SIMULATION SETUPS

#### A. Samples and Experimental Setups

Four samples were tested in this study: RT_1, RT_2, RTL_1, and RTL_2, as shown in Fig. 3(a). RT_1 and RTL_1 were subjected to 2 J impact energy, while RT_2 and RTL_2 experienced 6 J impact energy. The main distinction between these samples lies in the aging process: RT_1 and RT_2 were aged for one month in a salt spray chamber at room temperature, whereas RTL_1 and RTL_2 were unaged. All samples were composed of a PLA80%-PBAT20% matrix reinforced with flax fibers. For the polymer film production, the A500 matrix (PLA/PBAT 80/20 loaded with 10 wt% $CaCO_3$) was pre-dried at 70 °C for four hours. Films were extruded using a thermal profile along the screw, with temperatures set at 165 °C, 175 °C, 180 °C, 170 °C, and 165 °C (from the hopper to the die head), and a screw speed of 80 rpm. The resulting films had a thickness of approximately 100 microns.

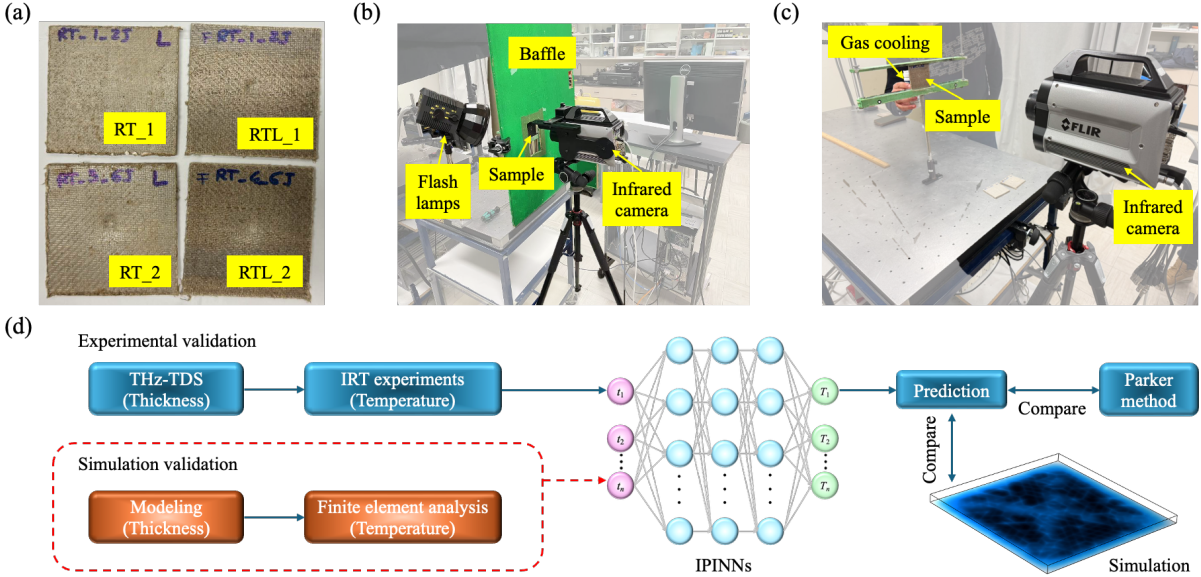

**Fig. 3.** Experiments and simulations for the thermal diffusivity measurement: (a) Samples. (b) Experimental setup for laser flash method. (c) Experimental setup for evaporative cryocooling method. (d) Training and prediction for inverse physics-informed neural networks (IPINNs).

To fabricate the laminates, samples were prepared by alternately stacking pre-dried A500 films (pre-dried at 70 °C for four hours) and flax fiber reinforcements (200 g/m²). The assembly, consisting of eight layers of flax fiber, was consolidated under the following conditions: processing temperature of 180 °C, and a stepwise pressure profile of 1, 5, 10, 15, 20, 25, and 30 bar (each maintained for 2 minutes). Final cooling was performed at ambient temperature under 40 bar pressure.

Two experimental methods were designed in this work, including the photothermal technique and the evaporative cryocooling method, as shown in Fig. 3(b) and (c). A cooled infrared camera (FLIR X8501sc, 3-5 µm, InSb, NEdT < 20 mK, 1280 × 1024 pixels) for both laser flash and evaporative cryocooling experiments was used. Two Xenon flash lamps (Balcar, 6.4 kJ for each, 2 ms) have been used for photothermal experiments.

### B. Simulation and Training Setup

The data hungry nature of deep learning significantly impedes the advancement of "AI for Science". To address this challenge, several researchers [36,37] have explored the use simulated data for training, achieving promising results. Fortunately, both PINNs and IPINNs adopt an unsupervised learning paradigm, which alleviates the need for large datasets. To validate the feasibility and accuracy of the proposed IPINNs framework combined with evaporative cryocooling excitation, both experiments and numerical simulations were conducted. The schematic of the training and prediction pipeline is illustrated in Fig. 3(d). The material properties used in the simulations are summarized in Table I. All training datasets were normalized (temperature values only; spatial coordinates remained unnormalized). Model implementation was carried out using the PyTorch framework and training was performed on NVIDIA 4060 Titan GPUs. It is worth noting that, due to extremely small magnitude of diffusivity is extremely small, the initial value of thermal diffusivity is set to $1 \times 10^{-8}$ m²/s.

TABLE I
The Material Properties for Simulation.

| Material | CFRP | GFRP | Steel | Wood | Concrete |
|---|---|---|---|---|---|
| Heat Conductivity: W/(m·K) | 0.651 | 0.3 | 50 | 0.12 | 1.4 |
| Density: kg/m³ | 1800 | 1800 | 7850 | 600 | 2400 |
| Heat Capacity: J/(kg·K) | 512.91 | 800 | 500 | 1700 | 880 |
| Thermal Diffusivity: mm²/s | 0.705 | 0.208 | 12.7 | 0.118 | 0.663 |

The neural network architecture consists of 2 neurons in the input layer, 8 hidden layers, each with 20 neurons, and a single neuron in the output layer. The activation function used is the Tanh function. During training, the Adam optimizer with a learning rate of $1 \times 10^{-4}$ is employed for 4000 iterations, supported by the ReduceLROnPlateau scheduler, which dynamically adjusts the learning rate to improve convergence. The loss function consists of three components: data fitting loss, physics-informed PDE residual loss (with an adaptive weighting factor $\lambda = \min(1.0, \frac{loss\_data}{loss_{pde} + 1 \times 10^{-6}})$), and boundary loss (weighted by 0.1). Gradient clipping with a maximum norm of 1.0 is applied to prevent gradient explosion.

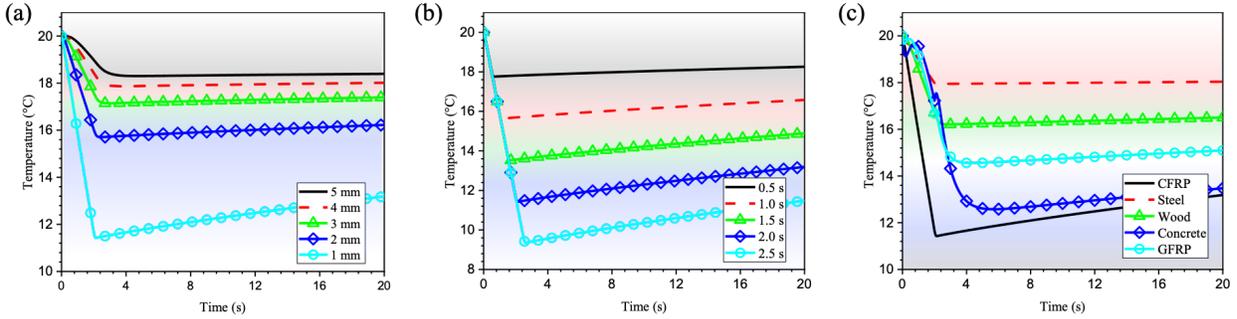

**Fig. 4.** Simulation results: (a) Samples (CFRP) with different thickness (1 mm - 5 mm). (b) Samples (CFRP) with different cooling time (0.5 s – 2.5 s). (c) Samples with different material properties.

The simulation results are shown in Fig. 4. It is evident that the amplitude and time delay (the time at which the temperature reaches its minimum) vary with sample thickness (Fig. 4(a)), cooling time (Fig. 4(b)), and material properties (Fig. 4(c)). Regarding sample thickness, the amplitude difference $\Delta T$ (where $\Delta T = |T_{min} - T_0|$, $T_{min}$ denotes the minimum temperature during cooling, $T_0$ denotes the initial temperature) decreases as the sample thickness increases, as shown in Fig. 4(a). Both the amplitude difference and time delay increase with the cooling time, as shown in Fig. 4(b). However, no clear positive or negative correlation between thermal diffusivity and amplitude / time delay, as shown in Fig. 4(c). This suggests that thermal diffusivity is not the sole factor determining the temperature profile, despite being the only constant in the isotropic one-dimensional heat conduction equation. Therefore, it is necessary to incorporate deep neural networks to address this nonlinear issue.

## IV. RESULTS AND DISCUSSION

### A. Experimental Results

Fig. 5(a) exhibits the thermograms obtained after principal component analysis (PCA) processing. Obviously, the impact resistance of samples (RT_1 and RT_2) decreases following salt spray aging. This degradation is attributed to the fact that salt spray aging deteriorates the mechanical properties of the polymer matrix, rendering the material more brittle and thus reducing its impact resistance. To further substantiate this observation, THz-TDS was employed, as shown in Fig. 5(b). The results reveal that the damaged area of RT_2 is noticeably larger than that of RTL_2.

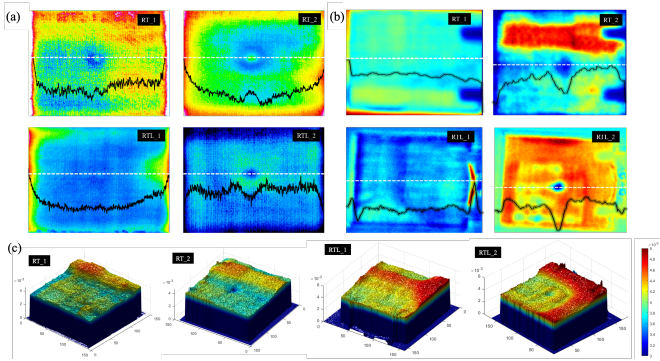

**Fig. 5.** Experimental results: (a) Thermograms after PCA processing. (b) THz-TDS images. (c) Thickness map of different samples based on THz-TDS.

In previous studies, the thermal diffusivity of impacted samples was typically calculated directly [38,39], without accounting for thickness variations resulting from impact damage. In this work, to more accurately characterize changes in thermal diffusivity, THz-TDS was first used to measure the thickness of each sample, as described by Eq. (18). As illustrated in Fig. 5(c), low-energy impacts (2 J) did not cause any permanent deformation in the samples (RT_1 and RTL_1). In contrast, a 6 J impact produced significant deformation at the center of the samples (RT_2 and RTL_2). Additionally, the deformation observed in RT_2 is greater than that in RTL_2, further supporting the earlier observation regarding the effects of salt spray aging.

After applying a simple affine transformation to align the thickness map with the thermograms, the real thermal diffusivity can be calculated using Eqs. (4) - (6). The results of samples RT_1 and RT_2 are presented in Fig. 6(a) and (b). It can be observed that the thermal diffusivity in the undamaged areas is approximately $4 \times 10^{-7}$ m²/s. Notably, the thermal diffusivity increases in the impacted area, with the peak value at the impact center reaching up to $6 \times 10^{-7}$ m²/s. This finding contrasts with the experimental results in Ref. [5]. In our study, the low-energy impact (6 J) leads to matrix densification, improved interfacial bonding between PLA and PBAT, and better fiber alignment, thereby enhancing thermal diffusivity. In contrast, the CFRP materials examined in Ref. [5] (thermal diffusivity map) and Ref. [39] (X-ray and photothermal tomograms) experienced high-energy impacts (20 J), which caused microcracks, fiber breakage, and interfacial debonding – ultimately leading to a decrease in thermal diffusivity. These results indicate that thermal diffusivity is closely related to the extent and nature of material damage. In particular, the observed increase or decrease in thermal diffusivity highlights its potential as an indicator for identifying internal failure modes such as matrix densification, cracking, and delamination.

The thermal diffusivity of RTL_1 and RTL_2 is shown in Fig. 6(c) and (d), where "mean" indicates the mean value from Eqs. (4) - (6). According to the results, Eq. (6) appears to be less sensitive to actual thermal diffusivity variations compared with Eqs. (4) and (5), as it shows no significant change at the center of the impacted area. In the undamaged area, the thermal diffusivity is approximately $4\times10^{-7}$ m²/s, while in the impacted area, it increases slightly to about $4.5\times10^{-7}$ m²/s. Compared to RT_2, RTL_2 exhibits a lower thermal diffusivity in the impact area, indicating a higher resistance to impact-induced structural changes.

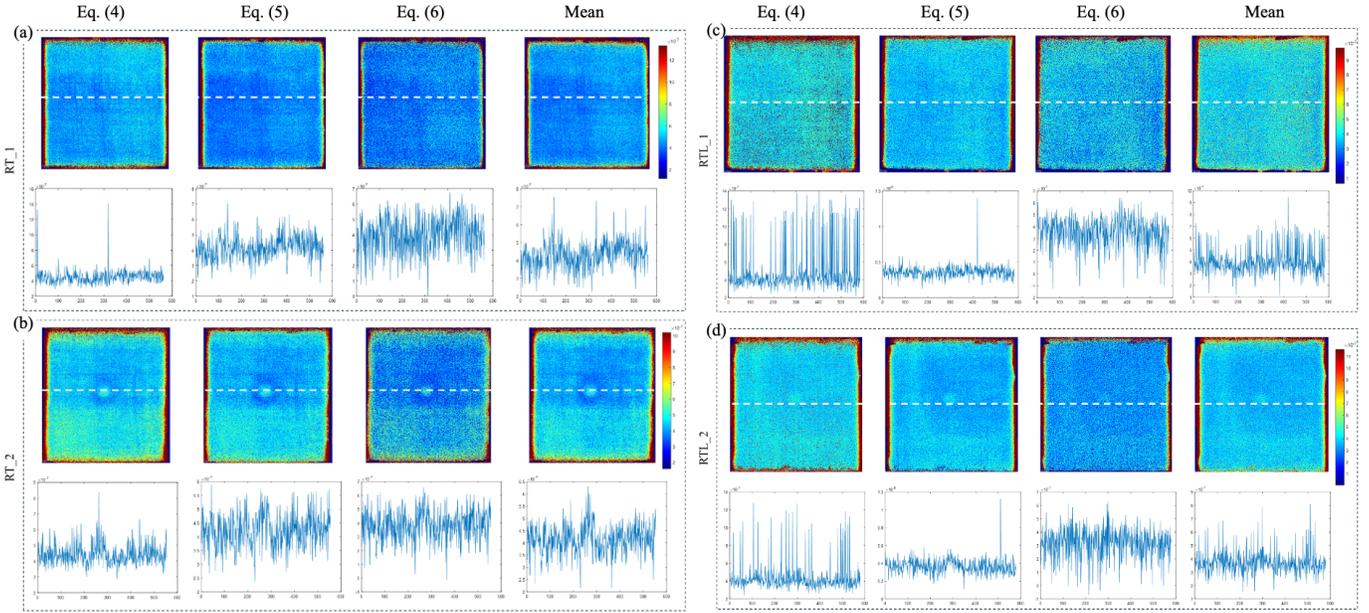

**Fig. 6.** Thermal diffusivity map of (a) RT_1, (b) RT_2, (c) RTL_1, (d) RTL_2 based on laser flash method and flash lamps.

### B. Simulation Validation for IPINNs

To evaluate the feasibility and accuracy of the proposed IPINNs combined with evaporative cryocooling method, numerical simulations were first conducted. The prediction results and corresponding loss curves are illustrated in Fig. 7. As discussed in the previous Section (Fig. 6), thermal diffusivity measurement is affected by multiple factors, including excitation (cooling) time and material's thickness. Conventional empirical models are inadequate for capturing the complexity of this multivariate relationship. Therefore, IPINNs were adopted in this study to address the challenge effectively.

As shown in Fig. 7(a), different types of materials were tested and trained using IPINNs framework, with their properties listed in Table I. Fig. 7(b) presents the results for varying cooling durations applied to a 3 mm thick CFRP plate, while Fig. 7(c) shows the training and prediction outcomes for samples of different thicknesses. The results indicate that as the number of training epochs increases, the loss value consistently decreases, demonstrating clear convergence of the training process. Additionally, the smooth loss curves reflect the stability of the entire training procedure. By comparing the predicted results with the ground truth, it is evident that the neural networks successfully capture the underlying features of transient heat conduction.

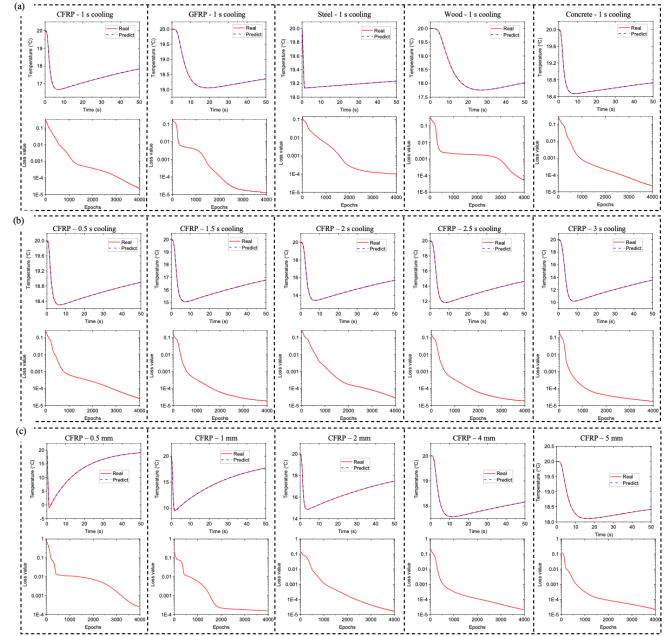

**Fig. 7.** Predicted temperature and loss value: (a) Comparing with different materials. (b) Comparing with different cooling time. (c) Comparing with different thickness. The sample's thickness in (a) and (b) is 3 mm, and the cooling time in (c) is 1 s.

Table II shows the thermal diffusivity results predicted by IPINNs. For different materials, the relative error between the predicted and actual values remains below 25%. When varying the excitation (cooling) time (0.5, 1.5, 2, 2.5, and 3 s), the relative error is less than 12%, with the minimum error reaching as low as 1.28%. For samples with varying thicknesses (0.5, 1, 2, 4, and 5 mm), the relative error remains under 5%, suggesting that sample thickness has minimal impact on IPINNs prediction accuracy. A relatively large error is observed when training on the steel sample. This is mainly attributed to the fact that the actual thermal diffusivity of steel is significantly higher than the initial value set during training. When the initial value of thermal diffusivity is adjusted to $1 \times 10^{-6}$ m²/s, the predicted result becomes $1.15 \times 10^{-6}$ m²/s, reducing the relative error to 9.45%. Therefore, it is essential to initialize the thermal diffusivity within a reasonable range to ensure accurate prediction.

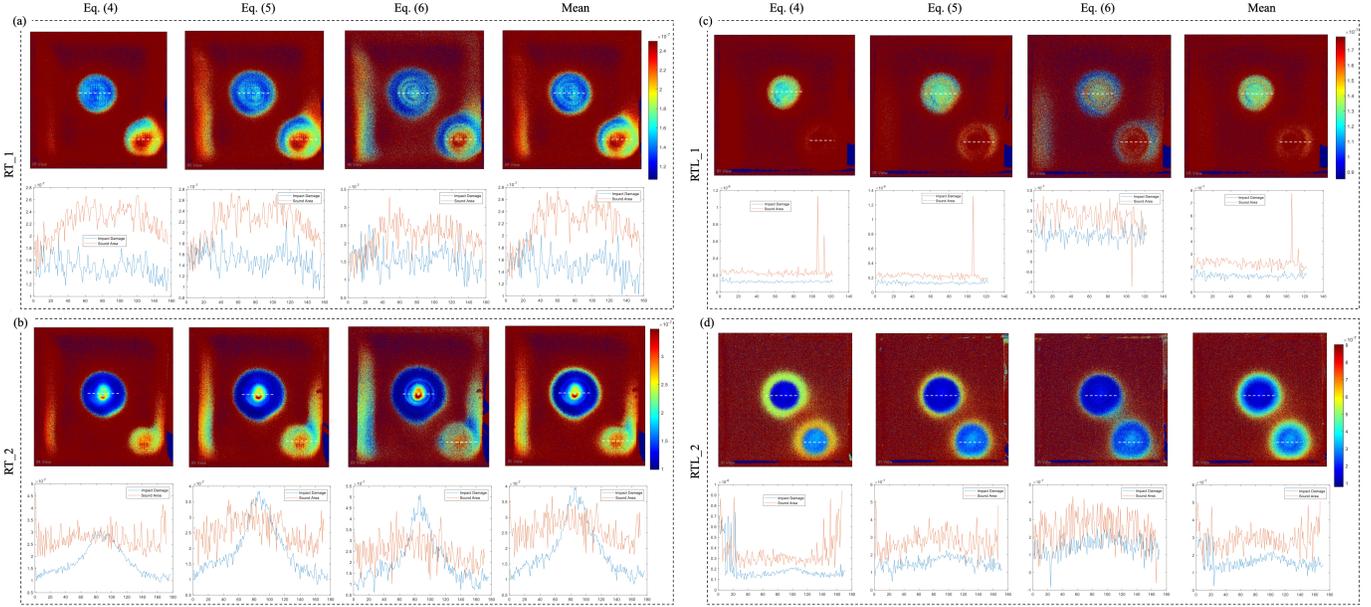

**Fig. 8.** Thermal diffusivity map of (a) RT_1, (b) RT_2, (c) RTL_1, and (d) RTL_2 based on laser flash method and evaporative cryocooling method.

TABLE II
The real and predicted thermal diffusivity (mm²/s).

| Parameter | IPINNs results | | | | |
|---|---|---|---|---|---|
| Material types | CFRP | GFRP | Steel | Wood | Concrete |
| Real | 0.705 | 0.208 | 12.7 | 0.118 | 0.663 |
| Prediction | 0.745 | 0.216 | 9.63 | 0.105 | 0.730 |
| Relative error | 5.70% | 3.85% | 24.17% | 11.01% | 8.60% |
| Cooling time | 0.5 s | 1.5 s | 2 s | 2.5 s | 3 s |
| Real | 0.705 | 0.705 | 0.705 | 0.705 | 0.705 |
| Prediction | 0.786 | 0.714 | 0.736 | 0.696 | 0.777 |
| Relative error | 11.49% | 1.28% | 4.40% | 1.28% | 10.21% |
| Material thickness | 0.5 mm | 1 mm | 2 mm | 4 mm | 5 mm |
| Real | 0.705 | 0.705 | 0.705 | 0.705 | 0.705 |
| Prediction | 0.730 | 0.713 | 0.716 | 0.699 | 0.736 |
| Relative error | 3.55% | 1.13% | 1.56% | 0.85% | 4.40% |

## C. Experimental Validation for IPINNs

Fig. 8 shows the thermal diffusivity maps obtained using the laser flash method. Unlike conventional photothermal technique, the temperature curves in evaporative cryocooling method exhibit an entirely opposite trend. To enable the application of the laser flash method, the original temperature data were inverted based on the initial temperature value.

By comparing the experimental results in Fig. 8 with those in Fig. 6, it is obvious that the laser flash method is not suitable for evaluating thermal diffusivity under evaporative cryocooling excitation. The primary reason is that the excitation (cooling) time used (2 s) is significantly longer than the 2 ms typically required by the method. As a result, neither theoretical formulas nor empirical coefficients can be reliably applied under evaporative cryocooling conditions. Although the authors attempted to shorten the cooling time, doing so led to minimal temperature variations, which are highly susceptible to noise from both the infrared camera and the surrounding environment.

Here, IPINNs were employed to estimate thermal diffusivity based on experimental data, which includes thickness measurements obtained from THz-TDS and temperature data collected via evaporative cryocooling method. The training configuration mirrors the setup used in the simulation. Fig. 9 shows the predicted results along with the corresponding loss curves. Due to the presence of unavoidable system and environmental noise in the experimental data, the loss values are higher than those obtained from the simulations. Nevertheless, the predicted results and the smooth loss curves demonstrate robust training performance of IPINNs in handling real experimental data.

Table III
The real and predicted thermal diffusivity (mm²/s).

| Parameter | IPINNs Results | | | |
|---|---|---|---|---|
| Material types | RT_1_D | RT_1_S | RT_2_D | RT_2_S |
| Real | 0.418 | 0.421 | 0.527 | 0.434 |
| Prediction | 0.3 | 0.3 | 0.455 | 0.311 |
| Relative Error | 23.23% | 28.74% | 13.66% | 28.34% |
| Material types | RTL_1_D | RTL_1_S | RTL_2_D | RTL_2_S |
| Real | 0.399 | 0.416 | 0.477 | 0.394 |
| Prediction | 0.4 | 0.396 | 0.642 | 0.501 |
| Relative Error | 0.25% | 4.81% | 34.59% | 27.16% |

*RT_1_D, RT_2_D, RTL_1_D, and RTL_2_D represent the samples in damage area. RT_1_S, RT_2_S, RTL_1_S, and RTL_2_S represent the samples in sound (non-damage) area.

Table III shows the predicted thermal diffusivity based on IPINNs. The actual thermal diffusivity values were obtained from the results in Fig. 6 after denoising and averaging. The maximum relative error is no more than 30%. Compared to the predicted thermal diffusivity from simulations, the experimental results exhibit a larger relative error. However, this error remains within an acceptable range when compared to the results shown in Fig. 8.

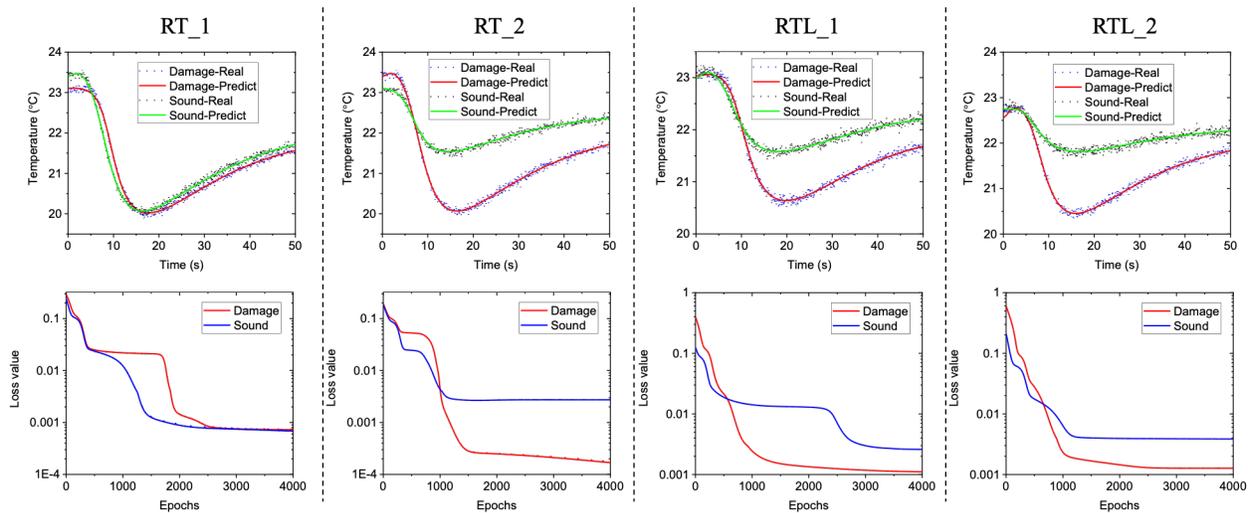

**Fig. 9.** Predicted temperature and loss value based on IPINNs and evaporative cryocooling.

Given that this is an ill-posed inverse problem, future work should consider increasing the number of observation points and accounting for lateral diffusion to further improve the accuracy of the IPINNs combined with the evaporative cryocooling method.

## V. Conclusion

In this study, a novel excitation modality, evaporative cryocooling, was introduced for measuring the thermal diffusivity of tested samples. To enhance the calculation accuracy of the conventional laser flash method, THz-TDS was employed to measure the samples' thickness. Observing the thermal diffusivity after low-velocity impact, it was found that low-energy impacts (2 J and 6 J) increase thermal diffusivity, while high-energy impacts lead to the opposite results (as shown in Ref. [5] and [37]). This is because low-energy impacts enhance matrix densification, improve PLA/PBAT interfacial bonding, and align fibers favorably, thereby increasing thermal diffusivity. In contrast, high-energy impacts induce microcracks, fiber breakage, and interfacial debonding, leading to a decrease in thermal diffusivity.

The evaporative cryocooling method cannot be classified as pulsed excitation, as the excitation (cooling) time significantly differs from the Dirac pulse (or the typical flash lamp duration around 2 ms). Therefore, conventional theoretical models and empirical formulas fail in this scenario. To address this, inverse physics-informed neural networks (IPINNs) were introduced and combined with the evaporative cryocooling method. Thermal diffusivity measurement is an ill-posed inverse problem with multiple influencing factors, such as sample thickness, cooling time, and excitation energy. To demonstrate the feasibility and accuracy of the IPINNs combined with the evaporative cryocooling method, both simulation and experiments were conducted and the results under various conditions were discussed. Both simulations and experiments confirmed the robustness of the proposed IPINNs approach. To further enhance measurement accuracy, future work should focus on increasing the number of observation points and constructing higher-dimensional IPINNs.

## Acknowledgment

This work is supported by the Natural Sciences and Engineering Research Council (NSERC) Canada through the Discovery and CREATE 'oN DuTy!' program (496439-2017), as well as the Canada Research Chair in Multipolar Infrared Vision (MiViM).

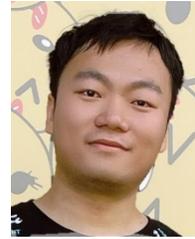

**Pengfei Zhu** (Graduate Student Member, IEEE) received the B.Eng. degree in engineering mechanics from North University of China, Taiyuan, China, in 2019, and the M.Eng. degree in solid mechanics from Ningbo University, Ningbo, China, in 2022. He is currently working toward the Ph.D. degree in electrical engineering with Université Laval, Québec, Canada.

His research interests include non-destructive testing, infrared thermography, deep learning, terahertz time-domain spectroscopy, and photothermal coherence tomography.

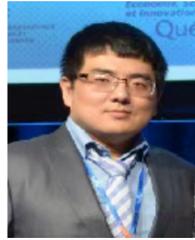

**Hai Zhang** (Member, IEEE) is a full professor at Harbin Institute of Technology, China. He received his BSc and MSc degrees from Shenyang University of Technology, China in 2004 and 2008, respectively, and his Ph.D. degree from Laval University, Canada, in 2017, where he is currently an adjunct professor. He was a Postdoctoral Research Fellow with University of Toronto, Canada. He was also a Visiting Researcher in Fraunhofer EZRT, Fraunhofer IZFP and Technical University of Munich, Germany.

His research interests include nondestructive testing, industrial inspection, medical imaging, infrared and terahertz spectroscopy, etc. He has authored or coauthored more than 150 technical papers in peer-reviewed journals and international conferences. He is also an Associate Editor for Infrared Physics and Technology, Measurement, and Quantitative InfraRed Thermography Journal, etc..

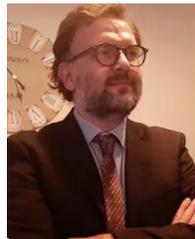

**Stefano Sfarra** received the Ph.D. degree in mechanical, management, and energy engineering from the University of L'Aquila (UNIVAQ), L'Aquila, Italy, in 2011. He worked as a Post-Doctoral Fellow at UNIVAQ until October 2017, where he became a Researcher in a fixed-term contract at the Department of Industrial and Information Engineering and Economics (DIIIE), UNIVAQ, Currently, he is an Associate Professor at DIIIE-UNIVAQ and an Adjunct Professor at Université Laval, Quebec, QC, Canada.

He is specialized in infrared thermography, heat transfer, speckle metrology, holographic interferometry, near-infrared reflectography, energy saving, and finite element simulation techniques. Concerning these research topics, he has published more than 200 papers in journals and international conferences.

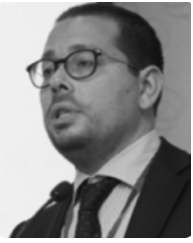

**Fabrizio Sarasini** received the Ph.D. degree in materials engineering from Sapienza University of Rome, Rome, Italy, in 2007. Since June 2016, he has been an Assistant Professor with the Faculty of Civil and Industrial Engineering, Sapienza University of Rome. He has more than 90 international peer-reviewed journal papers. His research interests include the impact response of composite materials for structural applications, the fiber/matrix interface adhesion, and the combination of natural fibers in hybrid composites.

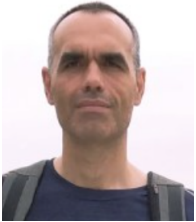

**Rubén Usamentiaga** (Senior Member, IEEE) received the M.S. and Ph.D. degrees in computer science from the University of Oviedo, Oviedo, Spain, in 1999 and 2005 respectively, with the Extraordinary Doctorate Award.
He is currently a Full Professor with the Department of Computer Science and Engineering, University of Oviedo. He was a Visiting Professor with the Universities of Laval and Pennsylvania. He has authored or coauthored more than eighty papers in JCR journals, with five Prize Paper Awards. In recent years, he has been working on several projects related to computer vision and industrial systems. His research interests include real-time imaging systems and thermographic observers for industrial processes. He is also a Senior Member of the IEEE and the IAS society.

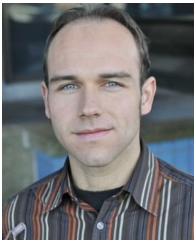

**Gunther Steenackers** was born in Vilvoorde, Belgium, in January 1977. He received the M.Sc. degree of electro-mechanical engineering and the Ph.D. degree from the Vrije Universiteit Brussel (VUB), Ixelles, Belgium, in July 2000 and 2008, respectively.
Since 2012, he has been a Full Professor with the University of Antwerp, Antwerp, Belgium, and a Guest Professor of Teaching Mechanics, Computer-Aided Engineering, and Finite Element Courses with VUB. He is also a member of different national and international consortia with a focus on establishing IR thermography as an optical measurement technique in an industrial context. In 2020, he became the Director of the Electromechanics Department, Faculty of Applied Sciences, UAntwerp. His current research focuses on IR thermography measurements, finite element modeling, and design optimization techniques.

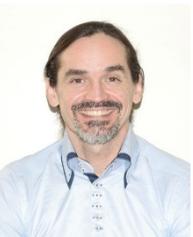

**Clemente Ibarra-Castanedo** received the M.Sc. degree in mechanical engineering (heat transfer) in 2000 from Université Laval, Quebec City, Canada, and the Ph.D. degree in electrical engineering (infrared thermography) in 2005 from the same institution. He is a professional researcher in the Computer Vision and Systems Laboratory at Université Laval and a member of the multipolar infrared vision Canada Research Chair (MIVIM). He has contributed to several research projects and publications in the field of infrared vision. His research interests are in signal processing and image analysis for the nondestructive characterization of materials by active/passive thermography, as well as near and short-wave infrared reflectography/transmittography imaging.

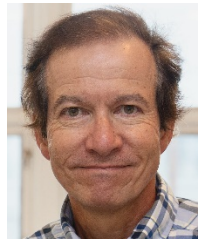

**Xavier Maldague** (Senior Member, IEEE) P.Eng., Ph.D. is full professor at the Department of Electrical and Computing Engineering, Université Laval, Québec City, Canada. He has trained over 50 graduate students (M.Sc. and Ph.D.) and contributed to over 400 publications. His research interests are in infrared thermography, NonDestructive Evaluation (NDE) techniques and vision / digital systems for industrial inspection. He is an honorary fellow of the Indian Society of Nondestructive Testing, fellow of the Canadian Engineering Institute, Canadian Institute for NonDestructive Evaluation, American Society of NonDestructive Testing. In 2019 he was bestowed a Doctor Honoris Causa in Infrared Thermography from University of Antwerp (Belguim).